\documentclass[11pt]{article}
\usepackage{jheppub}
\usepackage{amsmath}
\usepackage{amssymb}
\usepackage{braket}
\usepackage{bbold}
\usepackage{pifont}
\usepackage{color}

\newcommand{\beqn}{\begin{eqnarray}}
\newcommand{\eeqn}{\end{eqnarray}}
\newcommand{\dd}{\mathrm{d}}
\newcommand{\nn}{\nonumber}
\newcommand{\Tr}{\mathrm{Tr}}

\newcommand{\gmn}{g_{\mu\nu}}

\newcommand{\fmn}{f_{\mu\nu}}

\title{Ghost-free infinite derivative gravity}
\author{Brage Gording \& Angnis~Schmidt-May}
\affiliation{Max-Planck-Institut f\"ur Physik (Werner-Heisenberg-Institut)\\
F\"ohringer Ring 6, 80805 Munich, Germany}
\emailAdd{brageg@mppmu.mpg.de, angnissm@mpp.mpg.de}

\abstract{We present the construction of a gravitational 
action including an infinite series of higher derivative terms. 
The outcome is a classically consistent completion of a well-studied quadratic curvature theory.
The closed form for the full action is ghost-free bimetric theory, describing the interactions of a massive 
and a massless spin-2 field. At energies much smaller than the spin-2 mass scale, the theory reduces to 
general relativity. For energies comparable to the spin-2 mass, the higher derivative terms completing
the Einstein-Hilbert action capture the effects of the additional massive spin-2 field.
The theory is only ghost-free when the full series of higher derivatives is kept.}

\begin{document} 

\begin{flushright}
\hfill{MPP-2018-145} \vspace{20mm}
\end{flushright}
\maketitle
\flushbottom

\section{Introduction}

The Einstein-Hilbert action for general relativity (GR) is linear in the curvature $R_{\mu\nu}$
and thus corresponds to a two-derivative field theory. It describes the nonlinear self-interactions
of a massless spin-2 field. Adding higher derivative terms to the Einstein-Hilbert
action without introducing inconsistencies has been a long-standing problem. 
Such terms are expected 
to arise from quantum corrections in the effective field theory~\cite{tHooft:1974toh}.
They also appear in effective actions for string 
theory~\cite{Freeman:1986zh, Grisaru:1986vi, Gross:1986iv, Eliezer:1989cr}
and can possibly cure singularities of pure Einstein gravity~\cite{Siegel:2003vt}.

In this work we will consider the following 
four-derivative action proposed by Stelle~\cite{Stelle:1977ry},
\begin{align}\label{QCG}
S_{4}[g]=m_\mathrm{Pl}^2\int\dd^4x\,\sqrt{-g}
\Big[-2\Lambda + R
+\tfrac{1}{\mu^2}
\left(\tfrac{1}{3}R^2-R^{\mu\nu}R_{\mu\nu}\right)\Big]\,.
\end{align}
Its free parameters are the Planck mass $m_\mathrm{Pl}$, the cosmological constant $\Lambda$
and the mass scale~$\mu$.
The spectrum of the theory includes an additional massive spin-2 mode with mass~$\mu$
whose kinetic term in the action has a sign opposite to that of the massless mode.
This means that the action contains a spin-2 ghost giving rise to classical instabilities
since the energy is not bounded from below. The ghost is a direct consequence of the 
finite number of higher derivatives in the action, which is implied by Ostrogradsky's theorem~\cite{Ostrogradsky}.

In $d=3$ dimensions the above action can be rendered consistent by changing the overall
sign in front of the integral~\cite{Bergshoeff:2009hq}. This theory, called ``New Massive Gravity",
is ghost-free because in $d=3$ the massless spin-2 mode does not possess any local propagating degrees
of freedom. The procedure cannot be generalized to $d\geq 4$ where the massless mode
is dynamical~\cite{Ohta:2011rv, Kleinschmidt:2012rs}.\footnote{For an alternative approach 
of generalizing New Massive Gravity to any dimension (at the linearized level), see Ref.~\cite{Joung:2012sa}.}
In fact, the only ghost-free quadratic curvature theory in $d=4$ is given by $\int \dd^4x \sqrt{-g}\,R^2$
whose spectrum contains only a scalar in addition to the massless spin-2 
graviton~\cite{Alvarez-Gaume:2015rwa}.

Our goal in this work is to construct a ghost-free completion of the quadratic curvature 
action~(\ref{QCG}). Removing the ghost requires adding an infinite series of higher derivative terms
in order to avoid Ostrogradsky's theorem.
Our starting point will be ghost-free bimetric theory~\cite{Hassan:2011zd}, the only known
two-derivative theory describing interactions between a massless and a massive spin-2 field. 
Relations among bimetric theory and higher curvature gravity have been the subject of previous
investigations. In Ref.~\cite{Paulos:2012xe} a parameter scaling limit was taken 
which resulted in an auxiliary field formulation for the quadratic curvature action in (\ref{QCG}).
This construction required flipping the sign of a kinetic term which introduces the spin-2 ghost
into the action.
Ref.~\cite{Hassan:2013pca, Hassan:2015tba} instead derived a set of higher derivative
equations from the bimetric equations of motion. However, the higher derivative action obtained 
through the same procedure is not equivalent
to the original theory and hence its consistency is not guaranteed. 
Recently, Ref.~\cite{Akagi:2018osc} suggested that it could be still free from ghosts.
Lastly, Ref.~\cite{Cusin:2014zoa} derived an equivalent higher derivative action from bimetric
theory to lowest order in metric fluctuations around a flat space solution.

\vspace{3pt}

\paragraph{Summary of results.} 
In this work we outline the construction of a higher-derivative action for $\gmn$ 
whose lowest orders are given precisely by~(\ref{QCG}). We demonstrate that the resulting theory is 
classically equivalent to ghost-free bimetric theory with two tensor fields
$\gmn$ and $\fmn$, restricted to a wide class of solutions. 
Due to the on-shell equivalence, the untruncated higher curvature theory is free from ghosts at the
classical level. The higher order corrections are suppressed by the spin-2 mass scale of bimetric theory.
Bimetric theory also delivers a clear physical interpretation for the higher derivative terms: 
They describe the effects of a gravitating heavy spin-2 field with strong self-interactions.

\section{Ghost-free bimetric theory}
\subsection{The action for two tensor fields}
We begin by briefly reviewing the ghost-free bimetric theory for two symmetric tensor fields
$\gmn$ and $\fmn$ in vacuum. For more details we refer the reader to Ref.~\cite{Schmidt-May:2015vnx}.
The bimetric action is,
\beqn\label{bim}
S[g,f]=
S_\mathrm{kin}[g,f]
+S_\mathrm{int}[g,f]\,.
\eeqn
Here the Einstein-Hilbert kinetic terms are,
\beqn
S_\mathrm{kin}[g,f]=m_g^2\int\dd^4x\,\left(\sqrt{-g}\,R+\alpha^2\sqrt{-f}\,R^f\right)\,,
\eeqn
where $m_g$ and $\alpha m_g$ are the Planck masses and $R$ and $R^f$ are the Ricci 
scalars for the respective metrics.
The interaction potential is of the following form,
\beqn\label{intact}
S_\mathrm{int}[g, f]&=&
-2m^2m_g^2\int\dd^4x~\sqrt{-g}~\sum_{n=0}^4\beta_n\,e_n\big(S\big)\,,
\eeqn
which involves five free dimensionless parameters $\beta_n$ and a mass scale $m$. 
The interactions are given in terms of the square-root matrix 
$S^\mu_{~\nu}=(\sqrt{g^{-1}f}\,)^\mu_{~\nu}$ and 
the elementary symmetric polynomials $e_n(S)$ defined as,
\beqn
e_0(S)&=&1\,,\qquad
e_1(S)=\Tr\, S\,,\qquad
e_2(S)=\tfrac{1}{2}\big([\Tr\, S]^2-\Tr[S^2]\big)\,,\nn\\
e_3(S)&=&\tfrac{1}{6}\big([\Tr\, S]^3-3\Tr[S^2][\Tr\, S]+2\Tr[S^3]\Big)\,,
\qquad
e_4(S)=\det S\,.
\eeqn 
This specific structure of interactions is crucial for the consistency of the 
theory~\cite{deRham:2010kj, Hassan:2011hr, Hassan:2011zd}, since
it avoids the presence of a scalar mode known as the 
Boulware-Deser ghost~\cite{Boulware:1973my}. This scalar mode 
is an additional degree of freedom and therefore
different from the spin-2 ghost in quadratic curvature gravity.
Its presence in the action~(\ref{QCG}) is avoided by the choice of relative coefficient
$1/3$ between the two four-derivative terms.

\subsection{Bimetric mass spectrum}

The bimetric equations of motion admit maximally
symmetric solutions for which the two metrics are proportional,
$\fmn=c^2\gmn$. The proportionality constant $c\neq 0$ is determined by the following
polynomial equation~\cite{Hassan:2012wr},
\beqn\label{condc}
{\alpha^2}\left(c\beta_0+3c^2\beta_1+3c^3\beta_2+c^4\beta_3\right)
=\beta_1 +3c\beta_2+3c^2\beta_3+c^3\beta_4\,.
\eeqn
Around the proportional backgrounds, the equations for the linear metric fluctuations $\delta\gmn$ and 
$\delta\fmn$ can be decoupled by diagonalizing the mass matrix of the spin-2 modes.
The resulting spectrum consists of a massless and a massive spin-2 mode given by,
\beqn\label{modes}
\delta G_{\mu\nu}=\delta\gmn+\alpha^2\delta\fmn\,,\quad
\delta M_{\mu\nu}=\delta\fmn-c^2\delta\gmn\,.
\eeqn
The Fierz-Pauli mass of the massive fluctuation $\delta M_{\mu\nu}$ is,
\beqn\label{mass}
m_\mathrm{FP}^2=m^2\big(1+\alpha^{-2}c^{-2}\big)\big(c\beta_1+2c^2\beta_2+c^3 \beta_3\big)\,.
\eeqn
We emphasize that, in contrast to the quadratic higher curvature theory with similar spectrum, 
here both the massless and the massive spin-2 fields possess a healthy kinetic term
in the action. Bimetric theory does therefore not contain any ghost modes.

\subsection{Strong interaction limit}

In what follows we will focus on the parameter region where $\alpha\ll 1$.  
It is well-known that the exact limit $\alpha\rightarrow 0$ is the GR limit of bimetric theory 
when the metric $\gmn$ is coupled to matter~\cite{Baccetti:2012bk, Hassan:2014vja, Akrami:2015qga}.
We will now interpret this limit in terms of the spin-2 mass eigenstates.

Namely, for generic values of $\beta_n$ parameters, the Fierz-Pauli mass 
of the massive spin-2 mode becomes infinitely large in the 
$\alpha\rightarrow 0$ limit~\cite{Akrami:2015qga}. 
One can see this very easily by looking at simplifying parameter choices.
Consider for instance the model where all $\beta_n$ vanish except for $\beta_1$.
In this case eq.~(\ref{condc}) gives the solution $c^{-2}=3\alpha^2$. Reinserted into 
(\ref{mass}) we see that then indeed $m_\mathrm{FP}\rightarrow\infty$ for $\alpha\rightarrow 0$.
Another simple example is the case with $\beta_1=\beta_3=0$. Then eq.~(\ref{condc}) gives the solution
$c^2=\frac{3\beta_2-\alpha^2\beta_0}{3\alpha^2\beta_2-\beta_4}$, 
which approaches a constant in the limit
of vanishing $\alpha$. Again eq.~(\ref{mass}) then tells us that the mass goes to inifinity.
For more general parameters, one can simply think of solving (\ref{condc}) perturbatively in
$\alpha$. Then, at lowest order, the solution for $c$ is obtained by the vanishing of the
right-hand side of (\ref{condc}), which delivers the constant that $c$ approaches in the limit 
of $\alpha\rightarrow 0$. 
For later purposes it is useful to rewrite the equation (\ref{condc}) at lowest order as,
\beqn\label{ceqapprox}
s_0(c)\equiv \sum_{n=0}^3{3\choose n} c^{-n}\beta_{4-n}~=~0~+~\mathcal{O}(\alpha^2)\,.
\eeqn
Hence the Fierz-Pauli mass diverges for generic $\beta_n$ when $\alpha\rightarrow 0$ and
we conclude that this limit can be viewed as the limit of large spin-2 mass.

It was furthermore shown in a perturbative analysis in 
Ref.~\cite{Babichev:2016bxi} that the self-interactions of the massive spin-2 field $\delta M_{\mu\nu}$
become infinitely strong in the limit $\alpha\rightarrow 0$.
The self-interactions of the massless spin-2 mode $\delta G_{\mu\nu}$, 
on the other hand, are always precisely those of GR. 
They depend on the Planck scale and the cosmological constant but not on the value for $\alpha$. 
Hence they are insensitive to the strong interaction limit.

In the following we will use the bimetric equations of motion to integrate out the 
metric $\fmn$ in a perturbative setup with expansion parameter $\alpha\ll 1$. 
To lowest order in $\alpha$ this corresponds to neglecting all
dynamics of the massive, strongly self-interacting spin-2 field.
This procedure of integrating out $\fmn$ will result in an effective 
theory for the metric $\gmn$, whose fluctuations
$\delta\gmn\propto \delta G_{\mu\nu}-\alpha^2\delta M_{\mu\nu}$ are dominated by
the massless spin-2 mode for small values of $\alpha$. 
As expected, we will recover GR at low energies, that is at the zeroth order approximation in $\alpha$ 
where effects from the heavy spin-2 mode are completely ignored.
At higher energies, the effects of the massive spin-2 mode will enter the effective theory 
through higher curvature corrections to the Einstein-Hilbert term. 

We emphasize that the equations we solve are those for $\fmn$, whose linear fluctuations correspond 
to a superposition of the massless and massive spin-2 fields. This means that we will
integrate out part of the massless mode and hence the derivative
expansion that we obtain cannot really be viewed as a perturbative setup valid at low energies. 
Only the lowest order Einstein-Hilbert term gives a good approximation at low energies 
because the fluctuations of the metric $\gmn$ become massless in the limit $\alpha\rightarrow 0$.

\section{Construction of the infinite derivative theory}

\subsection{Outline of procedure}

Our aim is to solve the equations of motion for $\fmn$ which follow from varying the bimetric action (\ref{bim}),
\beqn\label{feq0}
\frac{\delta S[g,f]}{\delta f^{\mu\nu}}=0\,.
\eeqn
Our solution $\fmn(g)$ to these equations will be expressed in terms of $\gmn$,
its Christoffel connection and its curvatures.
Plugging this solution into the equations of motion for $\gmn$ 
results in a set of effective equations for $\gmn$ alone,
\beqn\label{effeq}
\left.\frac{\delta S[g,f]}{\delta g^{\mu\nu}}\right|_{f=f(g)}=0\,.
\eeqn
Alternatively, we can use the solution $\fmn(g)$ to remove $\fmn$ from the bimetric action
and obtain an effective action for $\gmn$ alone,
\beqn
\left.S_\mathrm{eff}[g]=S[g,f]\right|_{f=f(g)}
\eeqn
The equations for $\gmn$ following from $S_\mathrm{eff}[g]$ then are, 
\beqn\label{eqwint}
0=\frac{S_\mathrm{eff}[g]}{\delta g^{\mu\nu}(x)}=
\left.\frac{\delta S[g,f]}{\delta g^{\mu\nu}(x)}\right|_{f=f(g)}
+~\int\dd^4y~\frac{\delta f^{\rho\sigma}(y)}{\delta g^{\mu\nu}(x)} 
\frac{\delta S[g,f]}{\delta f^{\rho\sigma}(y)}
\,.
\eeqn
From this we see that when $\fmn$ is a solution to its own equation of motion~(\ref{feq0}),
then the equations for $\gmn$ obtained from $S_\mathrm{eff}[g]$ in (\ref{eqwint}) are equivalent to those
obtained from $S[g, f]$ in (\ref{effeq}). Note that we had to include an integral in (\ref{eqwint})
because $\fmn(g)$ will contain derivatives acting on $\gmn$ which upon integrating by parts
will act on ${\delta S}/{\delta f^{\rho\sigma}}$.
At this point we stress the difference between our approach and the one taken in 
Ref.~\cite{Hassan:2013pca, Hassan:2015tba}. The mentioned reference solved the $\gmn$ equations
(i.e.~${\delta S}/{\delta g^{\mu\nu}}=0$) for $\fmn$ and arrived at an
effective action which was not equivalent to the original one due to the additional operator acting on
${\delta S}/{\delta f^{\rho\sigma}}$. Here we will instead solve the $\fmn$ 
equations (\ref{feq0}) and obtain an effective theory for $\gmn$ which is fully equivalent 
to bimetric theory restricted to a wide class of solutions, as we will 
explain below.

Explicitly, the equations of motion for $\fmn$ read~\cite{Hassan:2011vm},
\beqn\label{feq}
\frac{\alpha^2}{m^2}\mathcal{G}^f_{\mu\nu}
=-f_{\mu\rho}\sum_{n=0}^3\beta_{4-n}(-1)^n\big[Y^{(n)}(S^{-1})\big]^\rho_{~\nu}\,,
\eeqn
where $\mathcal{G}^f_{\mu\nu}=R^f_{\mu\nu}-\frac{1}{2}\fmn R^f$ 
is the Einstein tensor and the contributions without derivatives 
coming from the potential are,
\beqn
\big[Y^{(n)}(S^{-1})\big]^\rho_{~\nu}=\sum_{k=0}^n(-1)^ke_k(S^{-1})[S^{k-n}]^\rho_{~\nu}\,.
\eeqn
Since the structure of the action is symmetric in both metrics, the equations for $\gmn$ 
have a very similar form. We will not need them in the following.

The equations (\ref{feq}) contain terms of two different orders in~$\alpha^2$. We will
solve them iteratively with an ansatz for $\fmn$ written as an expansion in $\alpha^2$. 
The lowest order is obtained from the potential contributions in (\ref{feq}) 
which are independent of $\alpha^2$. The next order is generated by plugging the lowest order 
into $\alpha^2\mathcal{G}^f_{\mu\nu}$ and thereby generating terms of order $\alpha^2$ which
again need to be cancelled by contributions from the potential. Iterating this procedure we can 
construct the solution to any order in $\alpha^2$. 
Our ansatz will therefore have the form of 
a perturbative expansion containing all possible covariant terms built from 
curvatures and covariant derivatives for~$\gmn$.

\subsection{Solution for $\fmn$}

In order to solve the $\fmn$ equations,
it is useful to start with an ansatz for the inverse square-root matrix 
$(S^{-1})^\mu_{~\nu}=(\sqrt{f^{-1}g}\,)^\mu_{~\nu}$. Including all possible
tensors up to order $\alpha^4$ we write,
\begin{align}\label{ansatzs}
(S^{-1})^\mu_{~\nu}&=a^{-1}\delta^\mu_{~\nu}
+\tfrac{\alpha^2}{m^2}\Big[b_1P^\mu_{~\nu}+b_2\Tr P\,\delta^\mu_{~\nu}\Big]
+\tfrac{\alpha^4}{m^4}\Big[c_1(P^2)^\mu_{~\nu}+c_2P^\mu_{~\nu}\Tr P
+c_3\Tr(P^2)\delta^\mu_{~\nu}
\nn\\&
+c_4(\Tr P)^2\delta^\mu_{~\nu}
+ c_5 \nabla^\mu\nabla_\nu P
+c_6 \nabla^2 P^\mu_{~\nu}
+c_7\nabla^2P\delta^\mu_{~\nu}+c_8g^{\mu\sigma}\nabla_\rho\nabla_{(\sigma} P^\rho_{~\nu)}
\Big]
+\mathcal{O}(\alpha^6)\,,
\end{align}
with coefficients $a$, $b_i$, $c_i$ to be determined from the equations.
The ansatz contains only tensors related to the metric $\gmn$ which is also used to raise
and lower indices.
Instead of using the Ricci curvature $R_{\mu\nu}$, we have parameterized the ansatz in terms
of the (rescaled) Schouten tensor, $P_{\mu\nu}=R_{\mu\nu}-\frac{1}{6}\gmn R$, for later convenience.
The terms of order $\alpha^4$ correspond to all possible tensors generated by plugging the 
order $\alpha^2$ terms into the curvatures for $\fmn$.
Note that we did not include the terms $\nabla^\mu\nabla^\rho P_{\rho\nu}$ and 
$\nabla^\rho\nabla^\sigma P_{\rho\sigma}\delta^\mu_{~\nu}$ 
since they are identical to other existing terms due to 
$\nabla^\rho P_{\rho\nu}=\nabla_\nu P^\alpha_{~\alpha}$
which follows from the Bianchi identity. 
We will comment on the generality of the ansatz below.

The coefficients in $(S^{-1})^\mu_{~\nu}$ are determined by inserting the ansatz (\ref{ansatzs}) into 
the equations (\ref{feq}) and comparing terms with the same order in $\alpha^2$.
From this we then obtain the solution for $\fmn$ by computing the inverse
$S^\mu_{~\nu}$ and using,
\beqn
\fmn = g_{\mu\rho}S^\rho_{~\sigma}S^\sigma_{~\nu}\,.
\eeqn
The calculation is somewhat lengthy but straightforward and 
conceptually identical to the one performed in Ref.~\cite{Hassan:2013pca}.
We present only the results here.

At lowest order the coefficient $a^{-1}$ in (\ref{ansatzs}) 
is constrained to satisfy the following polynomial equation,
\beqn\label{aeq}
\sum_{n=0}^3{3 \choose n}a^{-n}\beta_{4-n} =0\,.
\eeqn
The next two orders in $\fmn$ are then obtained as,
\begin{align}\label{pertsolf}
f_{\mu\nu} = a^{2}g_{\mu\nu}
-\tfrac{2\alpha^2}{m^2s_1}P_{\mu\nu}
&+\tfrac{\alpha^4}{m^4s_1^3a^2}\Big[ 
P_{\mu\rho}P^\rho_\nu\big(s_1-2s_2\big)
+2\text{Tr}(P)P_{\mu\nu} \big(s_2-s_1\big)
\nn\\
&
+\tfrac{1}{3}g_{\mu\nu}\text{Tr}(P^2)\big(2s_1+s_2\big)
+\tfrac{1}{3}g_{\mu\nu}\text{Tr}(P)^2\big(s_1-s_2\big)
\nn\\
&
-2s_1\Big(\nabla_\mu\nabla_\nu P +\nabla^2P_{\mu\nu}
-\nabla^\rho\nabla_\mu P_{\rho\nu}-\nabla^\rho\nabla_\nu P_{\rho\mu}\Big)\Big]\,,
\end{align}
where we have defined the constants,
\beqn
s_k&\equiv&\sum_{n=k}^3{3-k \choose n-k} a^{-n}\beta_{4-n}\,.
\eeqn
Equation (\ref{aeq}) which determines $a$ implies $s_0=0$. Since in general this equation is a third-order
polynomial in $a$, it will give rise to different branches of solutions. Note also that due to
the presence of infinitely many derivatives in the full solution for $\fmn$, its relation to 
the metric $\gmn$ is nonlocal. This is not surprising because solving the $\fmn$ equations 
requires inverting the derivative operators contained in the Einstein tensor.

Let us now briefly explain why the above ansatz cannot parameterize the 
most general solution to the bimetric equations. For a solution of the form (\ref{pertsolf}),
if $\gmn$ is a metric with constant curvature, $R_{\mu\nu}\propto \gmn$, then
$\fmn$ will necessarily be proportional to $\gmn$. 
Solutions that do not satisfy this property are known to exist and tend to show
pathological behaviour~\cite{Hassan:2014vja}. Our higher derivative theory for $\gmn$ will
not include these bimetric solutions since they have been eliminated by the ansatz (\ref{ansatzs}).
Nevertheless, the effective action for $\gmn$ will produce equations of motion that are identical to
the bimetric equations (\ref{effeq}). The only restriction is that we have solved the $\fmn$ equations
in a particular way, for instance, by choosing boundary conditions which exclude the pathological solutions.

\subsection{Higher derivative action}

\subsubsection{Kinetic and potential contributions}

Eliminating $\fmn$ from the kinetic terms is straightforward if one uses the relations,\footnote{We
use the conventions $[\nabla_\mu,\nabla_\nu]w_\rho=R_{\mu\nu\rho}^{\phantom{\mu\nu\rho}\sigma}w_\sigma$
and $R_{\mu\nu}=R_{\mu\rho\nu}^{\phantom{\mu\nu\rho}\rho}$.}
\beqn
R^f_{\mu\nu}&=&
R_{\mu\nu}-2\nabla_{[\mu}C_{\alpha]\nu}^{~~\alpha}
+2C_{\nu[\mu}^{~~\beta}C_{\alpha]\beta}^{~~\alpha}\,,
\nn\\
C_{\mu\nu}^{~~\alpha}&\equiv&\frac{1}{2}f^{\alpha\beta}\left(\nabla_\mu f_{\beta\nu}
+\nabla_\nu f_{\beta\mu}-\nabla_\beta f_{\mu\nu}\right)\,.
\eeqn
Inserting the expanded solution (\ref{pertsolf}) for $\fmn$, we can thus express 
the curvature $R^f$ for $\fmn$ in terms of curvatures 
$R_{\mu\nu}$ and connections $\nabla_\mu$ of $\gmn$ 
This allows us to write the Einstein-Hilbert term
for $\fmn$ entirely as an expansion in terms of $\gmn$. The result is,
\beqn\label{kinfcontr}
\alpha^2\sqrt{-f}\,R(f) = 
\sqrt{-g}\left[\alpha^2a^{2} R
+\frac{2\alpha^4}{s_1m^2}\left(R^{\mu\nu}R_{\mu\nu}-\frac{1}{3}R^2
+\nabla^2P^\alpha_{~\alpha}-\nabla^\mu\nabla^\nu P_{\mu\nu}
\right)\right]
+\mathcal{O}(\alpha^6)\,.\nn\\
\eeqn
The terms including covariant derivatives action on curvatures drop out
due to the Bianchi identity $\nabla^\nu P_{\mu\nu}=\nabla_\mu P^\alpha_{~\alpha}$.
Since the Einstein-Hilbert term for $\fmn$ already comes with a pre-factor 
$\alpha^2$ in the action,  obtaining its order $\alpha^{2n}$ contributions to the
effective action requires only orders up to $\alpha^{2n-2}$ in the expanded solution.
Interestingly, this also holds for the contributions from the potential.
In order to see this, we use the identity,
\beqn
\sqrt{-g}\sum_{n=0}^4\beta_ne_n(S)=\sqrt{-g}\sum_{n=0}^4\sigma_n e_n(M)\,,
\eeqn
with matrix $M^{\mu}_{~\nu}\equiv a^{-1}S^{\mu}_{~\nu}-\delta^{\mu}_{~\nu}$ whose perturbative
expression starts at first order in $\alpha^2$.
The identity involves the coefficients $\sigma_n$ defined as,
\beqn
\sigma_k&\equiv&\sum_{n=k}^4{4-k \choose n-k} a^{n}\beta_n\,.
\eeqn
Since $\sigma_1=0$ by equation~(\ref{aeq}), the term $e_1(M)=\Tr M$ in the potential
drops out from the action. Hence, the first nontrivial terms 
in the potential come from $e_2(M)$ which contains only quadratic terms.
It is thus sufficient to expand $M$ (or $S$) up to order $\alpha^{2n-2}$
in order to produce the $\alpha^{2n}$ contributions coming from the potential.
The contributions up to order $\alpha^4$ which require the solution expanded up to order
$\alpha^2$ read,
\beqn
\sqrt{-g}\sum_{n=0}^4\beta_ne_n(S) = 
\sqrt{-g}\left[\sigma_0+\frac{\alpha^4\sigma_2}{2m^4s_1^2a^4}
\left(\frac{1}{3}R^2-R^{\mu\nu}R_{\mu\nu}\right)\right] +\mathcal{O}(\alpha^6)\,.
\eeqn
Their structure is the same as that of the contributions (\ref{kinfcontr}) obtained from the kinetic term.

\subsubsection{Quadratic curvature terms}
By replacing $\fmn$ in all terms in the bimetric action we arrive at the final result
for the effective action up to order $\alpha^4$,
\begin{align}\label{final}
S_\mathrm{eff}[g]=\int\dd^4x\,\sqrt{-g}\left[m_\mathrm{Pl}^2\big(
 R-2\Lambda\big)
+\frac{\alpha^4c_{RR}}{m^2}
\left(\frac{1}{3}R^2-R^{\mu\nu}R_{\mu\nu}\right)\right]
+\mathcal{O}(\alpha^6)\,.
\end{align}
The coefficients are related to the original bimetric parameters through,
\beqn
m_\mathrm{Pl}^2= (1+a^{2}\alpha^2)m_g^2\,,
\qquad
\Lambda= \frac{\sigma_0m^2}{1+a^{2}\alpha^2}\,,
\qquad
c_{RR}= -\frac{1}{s_1}\left(2+\frac{\sigma_2}{a^4s_1}\right)m_g^2\,.
\eeqn 
All terms with derivatives acting on curvatures 
have dropped out from the action to this order in $\alpha$.
The ratio of the coefficients in front of the two lowest order terms is,
\beqn
\frac{m_\mathrm{Pl}^2m^2}{\alpha^4c_{RR}}
\propto \frac{1+a^{2}\alpha^2}{\alpha^{4}}s_1m^2
\propto (1+a^{-2}\alpha^{-2})\big(a\beta_1+2a^2\beta_2+a^3 \beta_3\big)m^2.
\eeqn
This is precisely the expression (\ref{mass}) for the Fierz-Pauli mass
if we replace $a\rightarrow c$. As we saw above, for $\alpha\ll 1$, 
the polynomial equation determining $c$ at lowest order is given by (\ref{ceqapprox}). 
This tells us that, to lowest order in $\alpha$, $c$ is the solution 
to the equation $s_0(c)=0$ and thus indeed coincides with $a$. 
We conclude that the curvature corrections in the effective action are suppressed 
by the spin-2 mass scale of bimetric theory.

\subsubsection{Cubic curvature terms}
Finally, we present the next order in the ghost-free completion of the quadratic curvature terms.
These are the six-derivative terms of order $\alpha^6$ in the effective action (\ref{final}) which 
can be brought into the form,
\begin{align}
S_6= \frac{\alpha^6m_g^2}{m^4}\int \text{d}^4x \,\sqrt{-g}\,&\Big(
u_1R^{\mu\rho}R_\rho^{~\nu}R_{\nu\mu}
+u_2R\,R^{\mu\nu}R_{\mu\nu}+u_3 R^3
\nn\\
&+u_4\big[2R^{\mu\nu}R^{\alpha\beta}R_{\mu\alpha\nu\beta}+R^{\mu\nu}\nabla^2R_{\mu\nu}-\tfrac{1}{3}R\nabla^2R\big]
\Big)\,.
\end{align}
Here, the coefficients $u_i$ are related to the bimetric parameters as follows,
\begin{align}
 u_1 &= \tfrac{2}{a^2s_1^3}\Big(-\tfrac{s_1}{2}+s_2+\tfrac{s_2\sigma_2
 -s_1\sigma_2}{a^4s_1}+\tfrac{\sigma_3}{3a^4}\Big)\,,
\qquad
u_2 = -\tfrac{2}{a^2s_1^3}\Big(\tfrac{5s_1}{4}+s_2+\tfrac{\sigma_2}{6a^4}
+\tfrac{s_2\sigma_2}{a^4s_1}+\tfrac{\sigma_3}{3a^4}\Big)\,,
\nn\\
u_3 &= \tfrac{1}{18a^2s_1^3}\Big(\tfrac{11s_1}{2}+7s_2-\tfrac{\sigma_2}{a^4}
+7\tfrac{s_2\sigma_2}{a^4s_1}+\tfrac{7\sigma_3}{3a^4}\Big)\,,
 \qquad
 u_4 = \tfrac{1}{a^2s_1^3}\Big(3s_1+\tfrac{2\sigma_2}{a^4}\Big)\,,
\end{align}
and we have dropped a total derivative term from the action.
We emphasize that the action expanded up to this order still propagates the spin-2 ghost
which can only be removed by the infinite number of derivative terms.

\section{Discussion}

We have presented the construction of an infinite derivative action for gravity 
which is classically equivalent to ghost-free bimetric theory
restricted to a wide class of solutions including the proportional backgrounds.
The expansion parameter is small when the spin-2 mass and interaction scale are large. 
The higher curvature terms in the effective theory for $\gmn$ 
thus encode the modifications of GR caused by a strongly self-interacting massive spin-2 mode.

The relative factor of ${1}/{3}$ between the quadratic curvature terms in 
(\ref{final}) reflects the absence of the scalar Boulware-Deser ghost in bimetric theory.
However, if the expansion is truncated at any finite derivative level, 
the incomplete theory has a spin-2 ghost instability. 
This ghost is removed by the infinite series of curvature corrections which contribute to the
propagator of the massive spin-2 field and change the sign of its residue. 
For small values of $\alpha$, it could in principle be possible to treat the higher-curvature corrections 
perturbatively, which requires carefully removing unphysical solutions containing
the instability. This can be done using a method proposed in Ref.~\cite{Simon:1990ic},
whose application to higher derivative corrections of the Einstein-Hilbert action 
was discussed in detail in Ref.~\cite{Parker:1993dk}. 
However, we do not expect this procedure to give a valid approximation of bimetric theory in our case.
The reason for this is that we have integrated out the metric $\fmn$ instead of the massive spin-2 mode.
This means that the higher derivative terms stem partly from expanding the propagator of the 
massless mode and hence diverge at low energies. Only the lowest order Einstein Hilbert term delivers
a good approximation for bimetric theory at energies much smaller than the Fierz-Pauli mass.

Our results here demonstrate that bimetric theory yields a ghost-free completion of
the well-studied curvature corrections. For practical applications it is better to work directly 
with bimetric theory where no unphysical solutions need to be isolated and where no infrared divergences occur.
Nevertheless, we believe that the structure of the higher derivative expansion 
is interesting from a theoretical point of view and can 
shed more light on earlier approaches taken in this direction.

For instance, higher orders in our expansion could be compared to the infinite derivative theory proposed
in Ref.~\cite{Biswas:2011ar}. They derived the most general quadratic curvature action
for $\gmn$ around a Minkowski vacuum with $R_{\mu\nu}=0$.  This type of expansion was 
considered prior in~\cite{Modesto:2011kw}. The actions are ghost-free and contain infinitely many derivatives
acting on the curvatures. They could thus be related to our action~(\ref{final}) expanded to second order
in curvatures around $R_{\mu\nu}=0$, but keeping all higher covariant derivatives. 
Verifying such a correspondence requires a better understanding of the
structure of higher orders in~(\ref{final}).

It is possible to make the two lowest order terms vanish in the effective action~(\ref{final})
by fixing $a^2=-\alpha^{-2}$ and $\sigma_0=0$. Together with $s_0=0$ this requires
bimetric parameters of the form
$
\beta_1=\beta_3=0$ and
$\beta_4=3\alpha^2\beta_2=\alpha^4\beta_0
$.
In this case the lowest order of the higher curvature theory
corresponds to the well-known conformal gravity action which is invariant under
Weyl transformations, $\gmn\mapsto \phi(x)\,\gmn$~\cite{Bach}. A similar feature already 
showed up in the analysis of Ref.~\cite{Hassan:2013pca, Hassan:2015tba} and the above
parameter choice is related to the presence of a partially massless spin-2 mode
in the linear spectrum and a constant scaling symmetry of the background~\cite{Hassan:2012gz}\,. 
Our result allows us to revisit the analysis of this particular bimetric model at the level of the higher 
derivative action. Due to the $\alpha$ dependence in the $\beta_n$ parameters, the perturbation theory 
needs to be revisited, which we leave for future work. 

We started from the bimetric action vacuum, but introducing matter coupled to
$\gmn$ is trivial. If matter was coupled to $\fmn$ instead, one would arrive at a highly modified
matter coupling in the effective theory for $\gmn$. 
We leave further investigations in this direction 
(which in spirit are similar to ideas pursued in Ref.~\cite{Luben:2018kll}) 
for future work. Moreover our analysis could be extended to general spacetime dimensions and
compared to the classifications of cubic curvature terms in Ref.~\cite{Oliva:2010zd}.
It would also be interesting to study the case of multiple heavy spin-2 modes 
and thereby continue investigations started in Ref.~\cite{Baldacchino:2016jsz}.

Finally, we believe that our result could be relevant for quantum gravity 
in the effective field theory approach. Quantum corrections to the Einstein-Hilbert action
are known to include terms of the form (\ref{final}) (see e.g.~Ref.~\cite{Calmet:2017rxl} and the
recent review \cite{Salvio:2018crh}) and it would be interesting to find out if there is a relation to the 
heavy spin-2 field of bimetric theory.
Such investigations could also help us to better understand how bimetric theory, 
as the ghost free completion of quadratic gravity, behaves during quantization.

\vspace{3pt}

\paragraph{Acknowledgements.} We are grateful to S.F.~Hassan, M.~L\"uben, J.~M\'endez-Zavaleta 
and M.~von Strauss for useful discussions and comments on the draft. 
This work is supported by a grant from the Max-Planck-Society.


\end{document}